# Crystals in the 19$^{th}$ century glass beads

Irina F. Kadikova,[1] Tatyana V. Yuryeva,[1] Ekaterina A. Morozova,[1] Irina A. Grigorieva,[2] Maria V. Lukashova,[3] Ilya B. Afanasyev,[4] Andrey A. Kudryavtsev,[3] and Vladimir A. Yuryev[5]

[1] *The State Research Institute for Restoration (GOSNIIR), Bldg. 1, 44 Gastello Street, Moscow 107014, Russia*

[2] *The State Hermitage Museum, 34 Dvortsovaya nab., Saint-Petersburg 190000, Russia*

[3] *TESCAN Ltd., PO Box 24, Office 212, 11 Grazhdansky Avenue, Saint-Petersburg 195220, Russia*

[4] *The Russian Federal Center of Forensic Science of the Ministry of Justice, Bldg. 2, 13 Khokhlovskiy Sidestreet, Moscow 109028, Russia*

[5] *A. M. Prokhorov General Physics Institute of the Russian Academy of Sciences, 38 Vavilov Street, Moscow 119991, Russia*

Corresponding author Irina F. Kadikova, e-mail: kadikovaif@gosniir.ru;
*Bldg. 1, 44 Gastello Street, Moscow 107014, Russia*
Tatyana V. Yuryeva, e-mail: yuryevatv@gosniir.ru
Ekaterina A. Morozova, e-mail: morozovaea@gosniir.ru
Irina A. Grigorieva, e-mail: grigorieva_ia@hermitage.ru
Maria V. Lukashova, e-mail: lukashovamv@tescan.ru
Ilya B. Afanasyev, e-mail: il.afanasyev@sudexpert.ru
Andrey A Kudryavtsev, e-mail: kudryavtsev.andrey@tescan.ru
Vladimir A. Yuryev, e-mail: vyuryev@kapella.gpi.ru

***Highlights***

- Thermal-fluctuation mechanism of corrosion is inherent to 19$^{th}$ century glass beads.
- 19$^{th}$ century glass beads contain crystals of different structure and sizes.
- The correlation “glass composition—a state of glass bead preservation” is found out.
- The research expands data about glass manufacture technology of the 19$^{th}$ century.

**Abstract**

Glass seed beads probably are the most numerous group of art historical glass. From the ancient times to the modern era, glass beads played an essential role in culture, as they were used, e.g., to decorate clothing, religious objects and functional tools, and served as an important good for global trade. The conservation state of items made of glass beads is an actual issue for curators and conservators. Glass, as well known, is often unstable material, so a number of internal and external factors are responsible for chemical and physical processes in glass and on the glass surface. It was noticed that some types of historical 19th century glass beads are subjected to more intense destruction than others. The samples of glass beads of different colours and conservation states were examined by means of SEM-EDX, EBSD, Raman and FTIR microspectroscopy.

Research showed that some types of beads are characterized by the presence of nano- and microcrystals in glass, which could be one of the causes of glass destruction. The elemental composition of crystals is different in each case and depends on raw materials and technical components used for glass manufacturing. The structure of crystals was identified by means of EBSD analysis.

Keywords: 19th century glass beads, crystallites in glass, glass deterioration, EBSD analysis

## 1. Introduction

Glass is one of the most important and mysterious material in the human history. From the ancient times to the modern era, it has been playing a major role in culture. Probably, glass seed beads are the most numerous group of glass historical items, as they were used, e.g., to decorate clothing, religious objects and functional tools, and served as an important good for global trade. The real golden age of beadwork art is related to the short period from the middle of the 18th century to the late 19th century when this craft became extremely fashionable and popular in Europe, in the North America, in Russia, etc. In this regard, the development and the beginning of the industrial-scale production of glass seed beads in the main glass-making centres in Murano (Italy) and Bohemia (in the present-day Czech Republic) [1] played a prominent role, producing a variety of glass beads diverged in size, colour and shape (round, faceted, multi-layered, transparent, cloudy, opaque, opalescent, etc.) [2]. It became possible due to various technological additives and new manufacturing technologies. So, at that time, the articles made of glass beads were an integral part of people everyday life and surrounded them at home, in the churches, etc., and now become a significant part of museums' collections in different countries and have a high historical context (*Fig. 1a*).

During the restoration process it was noticed that some kinds of beads are subjected to more intense destruction than others. For example, translucent turquoise ones change their colour to greenish, yellowish and black; a large network of cracks is observed on the transparent, red, peach and some other types of glass beads (*Fig. 1b-e*). As a result, the beads become very fragile and could not be restored and returned to their places on the embroidered issues. Surprisingly, the 19$^{th}$ century glass beads are destroyed to a much greater extent than even the ancient ones. So, the conservation state of items made of the 19$^{th}$ century glass beads is an actual problem for curators and conservators.

The article presents experimental results on technological examination of historical glass beads. In most of the samples micro inclusions of various sizes were detected in the glass matrix which could be one of the reasons of beads destruction [3, 4, 6]. The composition and crystal structure of these precipitates are identified and discussed.

## 2. Materials and methods

Samples of round and faceted historical glass beads of different colours and states of preservation were obtained from the museums' articles of the 19$^{th}$ century during restoration process. Also, the cross-sections of several turquoise glass beads with different states of deterioration were embedded into an epoxy matrix and grinded. It allowed us to analyse the elemental distribution (mapping) on the cross section area.

A complex of analytical methods was used to examine all the samples. MIRA 3 LMU (Tescan Orsay Holding) scanning electron microscope (SEM) with INCA-450 (Oxford Instruments Nanoanalysis) energy dispersive X-ray (EDX) spectrometer was used for imaging of glass structure, microanalysis and element mapping. Also, elemental composition of samples was analysed using a M4 TORNADO micro-X-ray fluorescence (micro-XRF) spectrometer (Bruker). It is equipped with a polycapillary lens, which enables focusing the tube radiation and concentrating it in a spot of about 25 μm in diameter. The X-ray detector is a silicon drift detector with a 30 mm$^2$ active area and energy resolution of about 145 eV. The presence of vacuum sample chamber permits identifying light elements that is very important for glass exploration, as Mg, Al, Si, K, and Ca are significant glass components and influence its properties.

MIRA 3 LMH (Tescan Orsay Holding) SEM equipped with Nordlys II S electron backscatter diffraction (EBSD) unit and AZtecHKL Advanced software (Oxford Instruments Nanoanalysis) was used for the structural analysis of crystallites in glass.

Besides, Libra-200 FE HR (Carl Zeiss) high-resolution transmission electron microscope (TEM) was also used to explore the structure of crystalline inclusions in glass of some samples.

In addition, vibrational microspectroscopy was used to study the crystals and glass. FTIR spectroscopy analysis was performed using a LUMOS microscope (Bruker) in attenuated total reflection (ATR) mode with Ge ATR crystal. Experiments were carried out in spectral range from 600 to 4000 $cm^{-1}$ at a spectral resolution of 4 $cm^{-1}$ (routine 64 scans were made).

Raman measurements were carried out using a Senterra confocal Raman microscope (Bruker) coupled to the Olympus BX51 microscope with 20× and 50× objectives. All spectra were acquired with a 1200 grooves/mm holographic grating, laser excitation wavelength 532 and 785 nm and power at the sample ranging from 0.2 to 10 mW.

Before all experiments, samples were washed with high purity isopropyl alcohol ([$C_3H_7OH$] > 99.8 wt. %) at 40°C for 20 minutes in a chemical glass placed into an ultrasonic bath (40 kHz, 120 W). Sample washing was repeated before every experiment.

## 3. Results and discussion

Stability of glass is well known to depend on a number of internal and external factors. Nowadays, the effect of external factors, especially humidity, on the destruction of glass is the basic theory on the glass corrosion, or glass disease, and is widely discussed in the professional literature [7, 8, 9, 10]. However, in our paper, we focus on the internal factors, such as the glass composition and the process of the glass beads manufacturing. We believe that these factors could cause physical and chemical processes in the glass and therefore in some cases make a major contribution in glass destruction. This is due to the observed patterns of destruction of some types of glass beads which are not characteristic for the glass disease. For example, usually corroded turquoise glass beads may be observed near the well-preserved ones in the embroidered article despite on they were subjected to the influence of humidity, organic pesticides, traces of handling, etc. equally (*Fig. 1c*).

One of the reasons of the glass beads destruction is related to its composition, and depends on the type and quality of raw materials used in its production. In the 19th century, the content of historical seed beads was very variable and included glass-forming components of natural origin, such as vein quartz, or another silicate raw material, which usually had different impurities (oxides of Al, Ca, Fe, Mg, etc.), potash from wood ashes, soda, calcium carbonate, lead-containing components, etc. To obtain the necessary characteristics of the artistic glass, various technological additives, such as fluxes, oxidizing agents, fining agents, colourizing, decolourizing or opacifying ones, were used [11]. Multicomponent content of the batch system and technological conditions of manufacturing processes of glass (first of all, high temperature) may cause to formation of new compounds. The research of glass beads of the 19th century showed that some kinds of beads both subjected to corrosion and well-preserved are characterized with the crystals in the glass matrix, which vary in size and content [3, 46]. The results of elemental analysis (SEM/EDX) of some samples of glass beads show the distinction in the ratio between the main components and technical additives in the beads of different colours (*Table 1*). The analysis of uncontrolled impurities in some historical glass bead samples was carried out by means of more sensitive micro X-ray fluorescence technique and discussed in our previous works [3].

Turquoise glass beads are subjected to more intense destruction than other ones, some of them had already become fragile, changed their colour significantly, and could not be returned to their place on the embroidery. So, there is a sufficient number of turquoise glass beads and their fragments (several tens of samples) for the very detailed research.

### 3.1. Turquoise glass beads

As a rule, potassium–sodium–lead glass ($K_2O$–$Na_2O$–PbO–$SiO_2$) with small content of Cu and Sb was used for manufacture of examined turquoise glass beads (*Table 1*, Samples 1–3). However, the composition of glass matrix is not homogeneous and depends on the state of glass preservation, i.e. it has been varying over time. For example, redistribution of Na and K is observed on the elemental map of the part of the deteriorated glass bead cross section (*Fig. 2c-e*) [6].

In the previous works, we found micro- (up to ~ 4 μm) and nanocrystallites of $KSbOSiO_4$ (KSS) (orthorhombic crystal system, $Pna2_1$ (33) space group; $a$ = 13.005 Å, $b$ = 6.475 Å, $c$ = 10.641 Å) in glass matrix (*Fig. 2a-f, 4a*), which nucleated during glass melting and bubbling [3, 4, 6]. Strong characteristic absorption bands of KSS were detected in the FTIR spectrum of glass bead (*Fig. 2f*), thus the high concentration of crystals in the glass matrix was revealed [12]. We concluded that large KSS precipitates and their clusters had given rise to internal stress (likely, tension) that caused strain (likely, tensile one) in the glass bulk. In course of time, it led to the conspicuous ‘internal corrosion’ of glass manifested in the gradual emergence of micro discontinuities (embryos of cracks) [13, 14, 15, 16] followed by the formation of a network of micro and then macro cracks in the glass bulk and slow fracturing of beads [3, 4].

Well-preserved glass beads (e.g., opal, yellow, red, two-layered-white ones, etc.) were not available to the research in a sufficient number, so in this case the results show features of the individual samples of glass beads.

### 3.2. Opal glass bead

A studied sample of faceted opal glass beads (Sample 4) was made of potassium–sodium glass ($K_2O$–$Na_2O$–$SiO_2$), yet the ratios between K and Na contents in opal and turquoise beads are different (Table 1). In the sample, Sb-rich particles were also found and identified as cubic $NaSbO_3$ crystals ($Fd\overline{3}m$ (227) space group; $a$ = 10.2 Å) with sizes ranging from several hundred nanometres to 10 microns (*Fig. 3a–c,*

*4b*). However, unlike turquoise beads, the opal ones are not susceptible to destruction with the formation of cracks.

### 3.3. Yellow glass beads

Stability of examined yellow glass beads relates to its composition – lead glass without alkali components ($PbO–SiO_2$). In some samples of yellow beads, Sb-rich crystals were detected (*Table 1*, Sample 5) (*Fig. 3d, e*) [4]. Its stability can also be explained by small crystallite sizes (~ 0.2 μm) and the fact that they do not form large clusters. We identified them as crystals of cubic lead antimonite, modified by iron atoms, $Pb_2Fe_{0.5}Sb_{1.5}O_{6.5}$ ($Fd\overline{3}m$ (227) space group; $a$ = 10.48 Å), which is a variety of Naples yellow (*Fig. 4c*).

The bands at 510, 450 and 300–350 cm$^{-1}$ in the Raman spectra can be used as diagnostic markers for the identification of the different types of lead antimonite [17]. The most intense line in the Raman spectra is peaked at about 138 cm$^{-1}$ and assigned to the Pb–O bond stretching vibrations (*Fig. 2g*). The line position depends on the content of structural units, firing temperature, and molar lead-to-antimony ratio. Differences in the spectra of the inclusions in the yellow bead and Naples yellow (Kremer, #10130) in the region 300–500 cm$^{-1}$ are due to variety of chemical contents (the presence of iron oxide in the crystal composition).

In another yellow glass bead (Sample 6), formation of crystals is not observed; however, the elemental composition of glass matrix is similar except for the component containing Sb (*Table 1*).

Interestingly, that in all the mentioned beads, Sb was detected only in the crystals and the impurity clouds surrounding them in the yellow beads [4] and was not found in the glass matrix. Thus, we conclude the high reactivity of Sb-contained components in glass melt.

### 3.4. Red glass bead

An examined sample of well-preserved red glass beads (Sample 7) was made of potassium–sodium–zinc glass ($K_2O–Na_2O–ZnO–SiO_2$) with small content of Ca (*Table 1*). The bright red colour of its glass is obtained through using colloidal staining, so we observe the uniform distribution of red particles (up to 1 μm) in this glass (*Fig. 3f–h*). Analysed their composition and structure by means of EDX and EBSD, we have identified the crystallites as hexagonal CdZnSSe ($P6_3mc$ (186) space group; $a$ = 4.052 Å, $b$ = 6.817 Å) [5].

### 3.5. Two-layered red-white glass bead

Glass compositions of the white inner layer and the red outer layer of two-layered red-white beads are similar (potassium–lead–sodium, $K_2O–PbO–Na_2O–SiO_2$); however, the ratios between Pb content is a little bit different (8.9 and 2.2 at. %, respectively) (*Table 1*, Sample 8) (*Fig. 3i, j*). In both layers, nanocrystals were detected. Crystals of $Ca_2Pb_3(AsO_4)_3Cl$ (hexagonal crystal system, $P6_3/m$ (176) space group; $a$ = 10.14 Å, $c$ =7.185 Å) (~ 200 to 300 nm) were identified in the white inner layer by means of EBSD. Crystals in the red outer layer have the similar elemental composition; however, we failed to obtain a good pattern of Kikuchi lines from them.

The state of preservation of the sample is rather good; fractures are only observed in red glass around the interface between the two layers. In this case, fracturing is related primarily with the manufacturing process of this kind of beads, i.e. with winding of the red glass melt on the solid white glass core. It followed by the vitrification of the former layer that resulted in emerging of the stressed domain at the interface. The uniform distribution of crystals, their nanoscale sizes, and the similar composition of glass in the both layers are probably the reasons of the good preservation of glass bead of this kind.

### 3.6. Glass beads with crystals of unidentified structure

SEM examination of a number of 19$^{th}$ century coloured glass beads showed that a lot of them have micro inclusions in the glass bulk. To the recent days, they were also detected in the peach (Sample 9, *Fig. 3k, l*), red-white (Sample 10, *Fig. 3m, n*), white-green (Sample 11, *Fig. 3o*), and light green (Sample 12, *Fig. 3p*) glass beads. As a rule, the size of the precipitates is very small, from several tens to several hundreds of nanometres, that makes their structure identification rather difficult. One of the important features of them is the uniform distribution in the glass matrix. Their detailed investigations and the identification the structure of these crystals by means of the EBSD technique are the goal of our upcoming researches.

### 3.7. Brief discussion

Based on the obtained results and the character of glass beads degradation, we propose a model of the destruction mechanism of the 19$^{th}$ century glass beads. According to our model, the internal stresses that occur in glass in the process of making beads contribute to its destruction most [3, 4, 1314, 15, 16]. Glass beads manufacturing is a very complicated multistage process. It consists of several stages at elevated temperature, including glass melting and bubbling followed by drawing, thinning and cutting of the glass tubes, tumbling, etc. After the melting, each of these stages includes annealing or even softening followed by cooling and the repeated vitrification. The cooling drastically affects the stability of glass [18]. However, even the compliance with technological requirements is not a key to success in durable glass bead production. For example, during tumbling, chopped pieces of a glass tube with a grinding mixture were processed at elevated temperatures (usually, no higher than 500°C), which led to the formation of superheated glass layer (i.e. modified by re-melting and re-vitrification) in outside and inner bead diameter, like a “shell” surrounding a “core”. This phenomenon is more characteristic for the low-fusible glasses [4]. As a result, the difference in density and in the coefficients of linear expansion of glass bulk and a layer of modified glass leads to internal stresses near their interfaces. Further destruction is caused by the process of gradual removal of the internal stress through the formation of cracks. Moreover, large crystalline inclusions and their clusters formed in the glass melt ($T > 1200$°C), due to the difference in coefficients of linear expansion of glass and crystals, might reinforce destruction of some types of glass beads during annealing. In the turquoise glass beads, this process is most intense in the modified glass layer (*Fig. 2c-e*). After the beads cooled down, all relatively fast processes cease; however, the significant internal stresses in glass matrix remain around crystals and/or at the core-shell interfaces, or at the interfaces between differently coloured glasses in the case of the two-layered beads. As a result, even at room temperature the slow process of nucleation and formation of micro cracks continues, cracks grow and form networks, increasing its fragility. Thus, glass beads destruction is a long-term process of relaxation of internal stress in glass through the nucleation and growth of cracks, which leads to the glass breakdown into individual relaxed grains.

This model may be understood by considering the fluctuation mechanism of the fracture of solids at the atomic level [15]. The process resulting in the fracture of a stressed solid is known to comprise a sequence of elementary events of breaking of loaded atomic bonds by local energy fluctuations. In internally stressed glass, the local stress lowers the atomic bond rupture barrier, reducing the energy fluctuation needed to break the bond that results in accelerated rupturing of the loaded bonds [13, 15]. As a result, micro discontinuities emerge in glass at locations of tensile, shear and rotational strains [14] that give rise to the glass fracturing process. It should be noted that the characteristic sizes of the arising micro discontinuities [14] coincide with the characteristic dimensions of light scatterers previously observed by us in glass of deteriorating turquoise beads – the latter were estimated as $\ll 150$ nm [19]. The number density of the light scatterers grows with the rising degree of glass deterioration. We presumably identify the observed light scatterers with the micro discontinuities [13]. The glass destruction related to the formation and growth of such microscopic defects is a kinetic thermal fluctuation process occurring throughout the entire period of material loading [14]. Their merger and formation of micro cracks is the only possible outcome of this process. Then, micro cracks elongate and widen, a network of macro cracks forms and a bead breaks.

## 4. Conclusion

The destruction of different types of 19$^{th}$ century glass beads is an important issue in preservation of embroidered objects in the museums’ collections. According to the modern conception, glass corrosion, or glass disease is considered one of a major factors explaining the destruction of glass. So, all the proposed conservation methods are focused on the protecting of glass surface from atmospheric moisture [9, 10]. However, the results of investigation of the representative group of unstable turquoise and some well-preserved coloured glass beads show unusual patterns of destruction, including generation of micro cracks in the glass bead core rather than on its surface, change of glass colour, formation of fragile modified glass layer in the inner surface, etc. Interestingly, the presence of nano- and microcrystals in glass matrix is a characteristic feature of both deteriorated and well-preserved beads. The elemental composition of crystals is different in each case and depends on raw materials and technical components used for glass manufacture. In the glass beads, the crystallites are the sources of the internal stresses and cause to glass destruction in more or less extent. The direst consequences of their influence on the state of glass preservation are observed on the turquoise glass beads where large KSS crystals and their clusters formed.

According to the results of the study, we suggest a model of destruction, which reflects the specific features, such as the composition and characteristics of manufacturing processes of glass beads. Based on the proposed mechanism, we believe that the deterioration of some types of beads is irreversible, thus the major task of the future research is to find the way to slow down the process of glass destruction.

**Acknowledgment**

We express our deep gratitude to the A.S. Pushkin State Museum and the Museum-Estate “Kuskovo” for providing samples (fragments of damaged beads). We also thank the leadership of the State Historical Museum and the Serpukhov Historical and Art Museum for the opportunity to get acquainted with the collections of objects made of glass beads.

We thank our research team member Miss Darya Klyuchnikova (GOSNIIR) for preparation of samples to experiments. We also grateful to Ms. Lyubov Pelgunova of A. N. Severtsov Institute of Ecology and Evolution of RAS for XRF analyses.

The research was accomplished under the non-profit collaboration agreement between GOSNIIR and GPI RAS and under the non-profit collaboration agreement between GOSNIIR and IGIC RAS.

**Funding**

The study was conducted with the financial support of the Russian Science Foundation, grant number 16-18-10366.

**References**

1. M.-J. Opper, Howard Opper. “French Beadmaking: An Historical Perspective Emphasizing the 19th and 20th Centuries,” *BEADS: Journal of the Society of Bead Researchers* 3 (1991) 44–59. https://surface.syr.edu/beads/vol3/iss1/5.
2. E. S. Yurova, “Epoch of Beads in Russia.” Moscow, Russia: Interbook-business, 2003 [In Russian].
3. T. V.Yuryeva, I. B. Afanasyev, E. A. Morozova, I. F. Kadikova, V. S. Popov, V. A. Yuryev. “$KSbOSiO_4$ Microcrystallites as a Source of Corrosion of Blue-Green Lead-Potassium Glass Beads of the 19th Century,” *Journal of Applied Physics* 121 (2017) 014902. https://doi.org/10.1063/1.4973576.
4. T. V. Yuryeva, , E. A. Morozova, I. F. Kadikova, O. V. Uvarov, I. B. Afanasyev, A. D. Yapryntsev, M. V. Lukashova, S. A. Malykhin, I. A. Grigorieva, V. A. Yuryev. “Microcrystals of Antimony Compounds in Lead-Potassium and Lead Glass and Their Effect on Glass Corrosion: A Study of Historical Glass Beads Using Electron Microscopy,” *Journal of Materials Science* 53 (2018) 10692–10717. https://doi.org/1007/s10853-018-2332-2.
5. T. V. Yuryeva, S. A. Malykhin, A. A. Kudryavtsev, I. B. Afanasyev, I. F. Kadikova, V. A. Yuryev, “CdZnSSe crystals synthesized in silicate glass: Structure, cathodoluminescence, band gap, discovery in historical glass, and possible applications in contemporary technology,” *Materials Research Bulletin* 123 (2020) 110704. https://doi.org/10.1016/j.materresbull.2019.110704.
6. I. F. Kadikova, I. A. Grigorieva, E. A. Morozova, T. V. Yuryeva, I. B. Afanasyev, and V. A. Yuryev. “A Comprehensive Study of Glass Seed Beads of the 19th Century Using Vibrational Spectroscopy and Electron Microscopy,” Poster presented at the XXVII Russian Conference on Electron Microscopy and the V School for Young Scientists “Modern Methods of Electron and Probe Microscopy in Studies of Organic, Inorganic Nanostructures and Nano-Biomaterials”, RCEM-2018, Chernogolovka, Moscow region, Russia, August 28–30, 2018 [In Russian]. https://doi.org/10.13140/RG.2.2.24263.96167.
7. R. H. Brill, “Crizzling – A Problem in Glass Conservation.” In *Conservation in Archaeology and the Applied Arts, Stockholm Congress*, edited by N. S. Brommelle and P. Smith, 121–134, London, UK: International Institute for Conservation of Historic and Artistic Works (IIC), 1975. https://doi.org/10.1179/sic.1975.s1.021.
8. J. J. Kunicki-Goldfinger, “Unstable Historic Glass: Symptoms, Causes, Mechanisms and Conservation,” *Studies in Conservation* 53(sup2) (2008.) 47–60. https://doi.org/10.1179/sic.2008.53.Supplement-2.47.

9. S. P. Koob, "Crizzling Glasses: Problems and Solutions," *Glass Technology – European Journal of Glass Science and Technology Part A* 53(5) (2012) 225–227. http://www.ingentaconnect.com/content/sgt/gta/2012/00000053/00000005/art00008.
10. R. O'Hern, K. McHugh, "Red, Blue, and Wound All Over: Evaluating Condition Change and Cleaning of Glass Disease on Beads." In *American Institute for Conservation of Historic and Artistic Works Objects Specialty Group Postprints, Volume 21. Proceedings of the Objects Specialty Group Session, 42nd Annual Meeting, San Francisco, California, Conscientious Conservation: Sustainable Choices in Collection Care, May 28–31, 2014,* edited by Suzan Davis, Kari Dodson, and Emily Hamilton, 2014, 205–228. http://resources.conservation-us.org/osg-postprints/postprints/v21/ohern.
11. V. A. Galibin, "*Composition* of *Glass* as an *Archaeological Source*: Ars Vitraria Experimentalis." In *Proceedings of the institute of the history of material culture, volume IV*. Saint Petersburg, Russia: Russian Academy of Sciences, 2001 [In Russian].
12. I. F. Kadikova, E. A. Morozova, I. A. Grigorieva, T. V. Yuryeva, V. A. Yuryev, "Infrared radiation absorption in crystals of orthorhombic KSbOSiO4: the determination of characteristic absorption bands for spectral analysis of glass and ceramics." Preprint, 2019. https://doi.org/10.13140/RG.2.2.19235.91682/3.
13. I. F. Kadikova, E. A. Morozova, T. V. Yuryeva, I. A. Grigorieva, V. A. Yuryev, "Glass depolymerization in the process of long-term corrosion: A study of deteriorating semiopaque turquoise glass beads using micro-FTIR spectroscopy," *Materials Research Express* 7 (2020) 025203. https://doi.org/10.1088/2053-1591/ab7510.
14. V. I. Betekhtin, A.G. Kadomtsev, 'Evolution of microscopic cracks and pores in solids under loading," *Physics of the Solid State* 47 (2005) 825–831. https://doi.org/10.1134/1.1924839.
15. A. I. Slutsker, "Atomic-level fluctuation mechanism of the fracture of solids (Computer simulation studies)," *Physics of the Solid State* 47 (2005) 801–811. https://doi.org/10.1134/1.1924836.
16. V. P. Pukh, L. G. Baikova, M. F. Kireenko, L. V. Tikhonova, T. P. Kazannikova, A. B. Sinani, "Atomic structure and strength of inorganic glasses," *Physics of the Solid State* 47 (2005) 876–881. https://doi.org/10.1134/1.1924836.
17. F. Rosi, V. Manuali, T. Grygar, P. Bezdicka, B. G. Brunetti, A. Sgamellotti, L. Burgio, C. Seccaroni, C. Miliani, "Raman scattering features of lead pyroantimonate compounds: implication for the non-invasive identification of yellow pigments on ancient ceramics. Part II. *In situ* characterisation of Renaissance plates by portable micro-Raman and XRF studies," *Journal of Raman Spectroscopy* 42 (2011) 407–414. https://doi.org/10.1002/jrs.2699.

18. I. I. Kitaygorodsky (Ed.), "Glass Technology." Moscow, USSR: The State Publishing House of Literature on Building, Architecture and Construction Materials, 3 edition 1961. http://www.klin-lab.ru/images/books/Kitaygorodskiy_Tehnologiya_stekla.djvu.
19. T. V. Yuryeva, V.A. Yuryev, "Degradation and destruction of historical blue-green glass beads: a study using microspectroscopy of light transmission," *Journal of Optics* 16 (2014) 055704. http://doi.org/10.1088/2040-8978/16/5/055704.

**Figures and Tables**

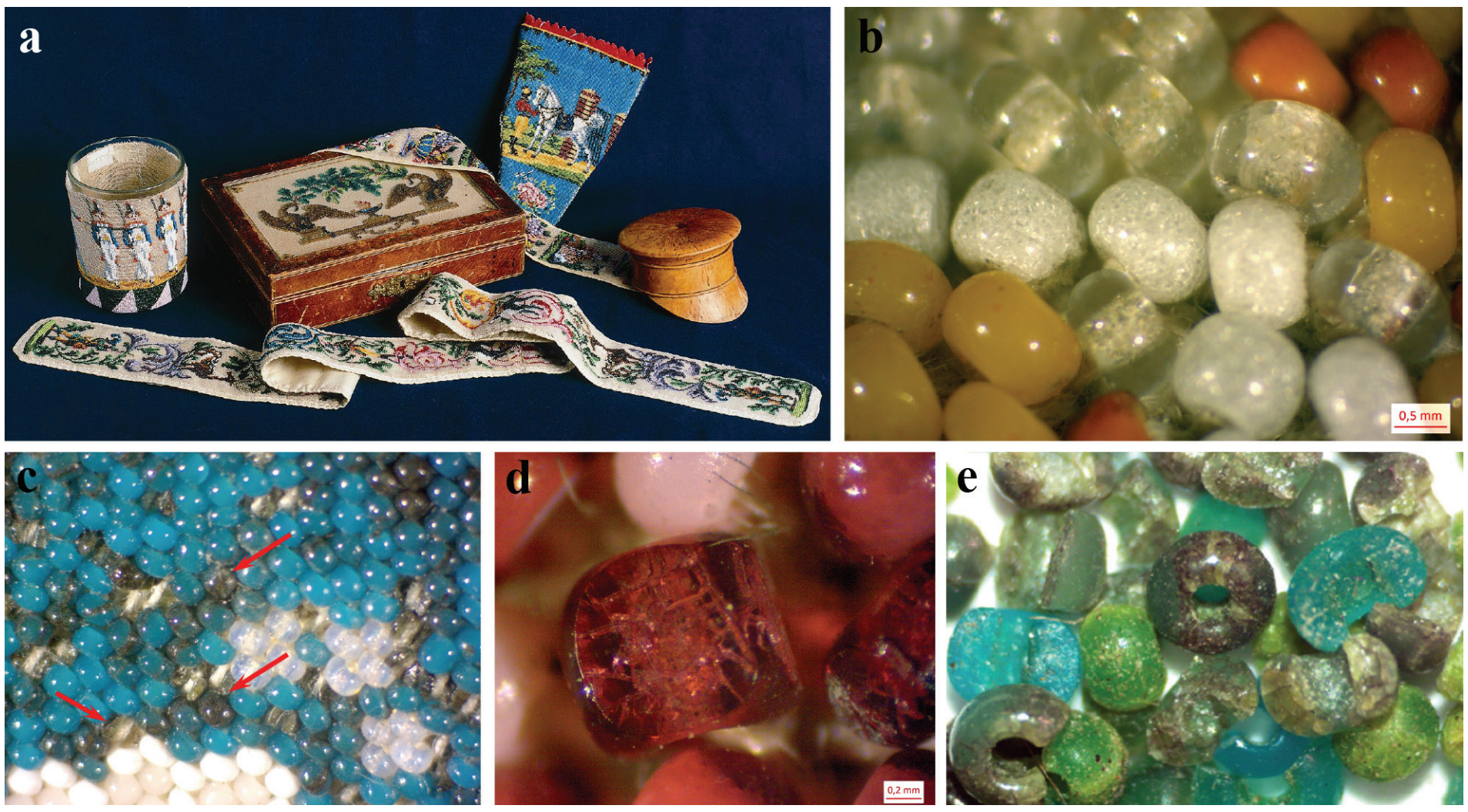

*Fig. 1.* Embroidered articles from the Russian museum collections:

(a) exhibits from '*Glass Beaded Passion, or Three Centuries of Beads in Russian Art'* exposition, the Ostankino Museum-Estate (Moscow, Russia);

(b–d) fragments of embroideries demonstrating corroded glass beads (the red arrows point some examples of destructed historical turquoise beads);

(e) deteriorated and mechanically damaged turquoise glass beads from museum exhibits of the 19$^{th}$ century.

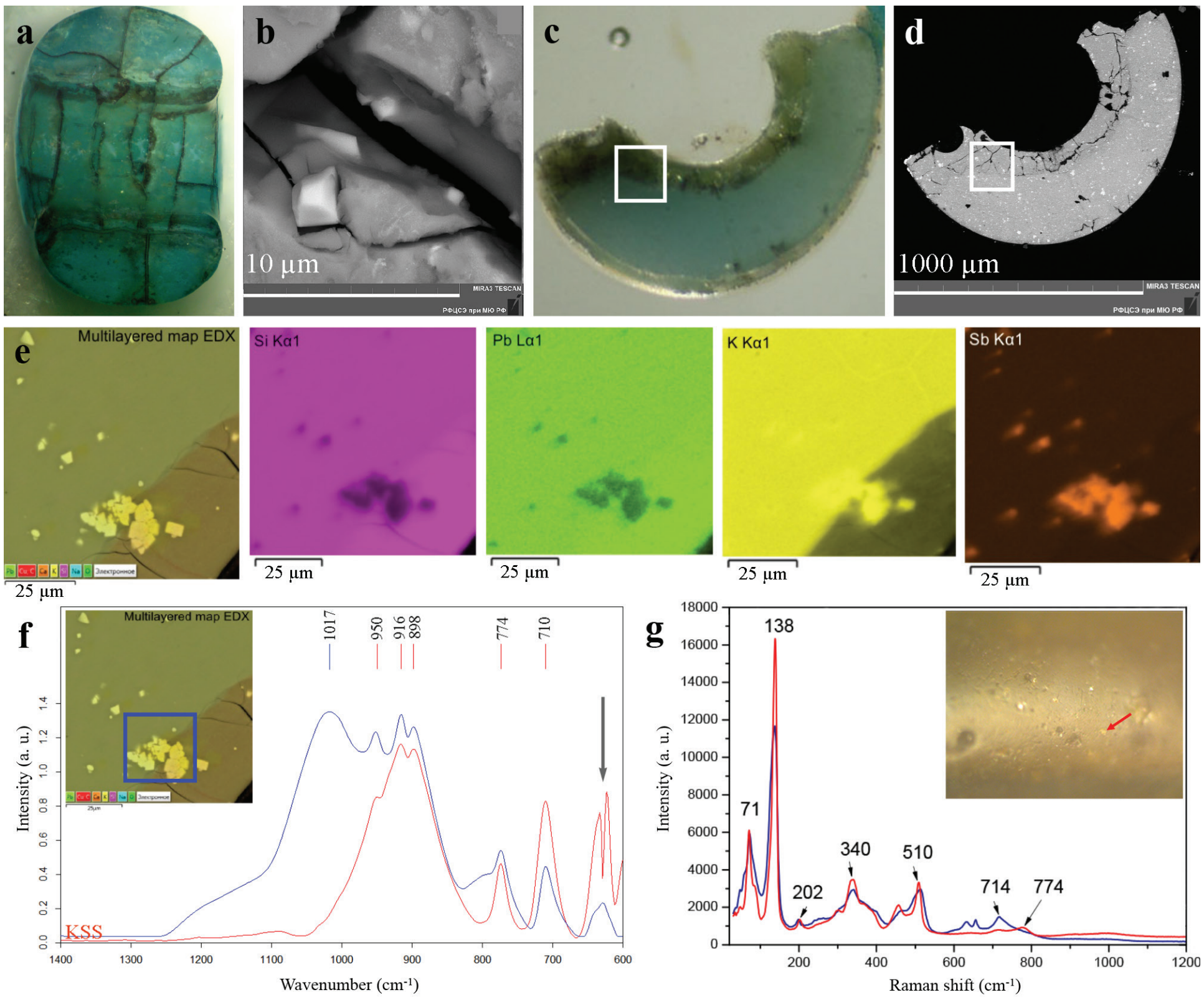

*Fig. 2.* A case study of micro inclusions in turquoise (Samples 1 and 3) and yellow (Sample 5) beads:
(a) a micro photograph of the typically destructed turquoise glass bead (Sample 1) in polarized light and
(b) a SEM image of a large $KSbOSiO_4$ crystal in the Sample 1;
(c) a micro photograph of the cross-section of the turquoise glass bead (Sample 3) in polarized light;
(d) a SEM image and (e) EDX elemental maps (multilayer, Si, Pb, K and Sb) of the cross-section domain shown by the white frame in the panel (d);
(f) FTIR spectra of the Sample 3 from the area (shown by the blue frame), in which $KSbOSiO_4$ crystals were identified; the vertical gray arrow indicates an artificial band splitting due to the instrument construction;
(g) Raman spectra of (1) yellow inclusion in the yellow bead pointed by red row (Sample 5) and (2) Naples yellow pigment (Kremer, # 10130).

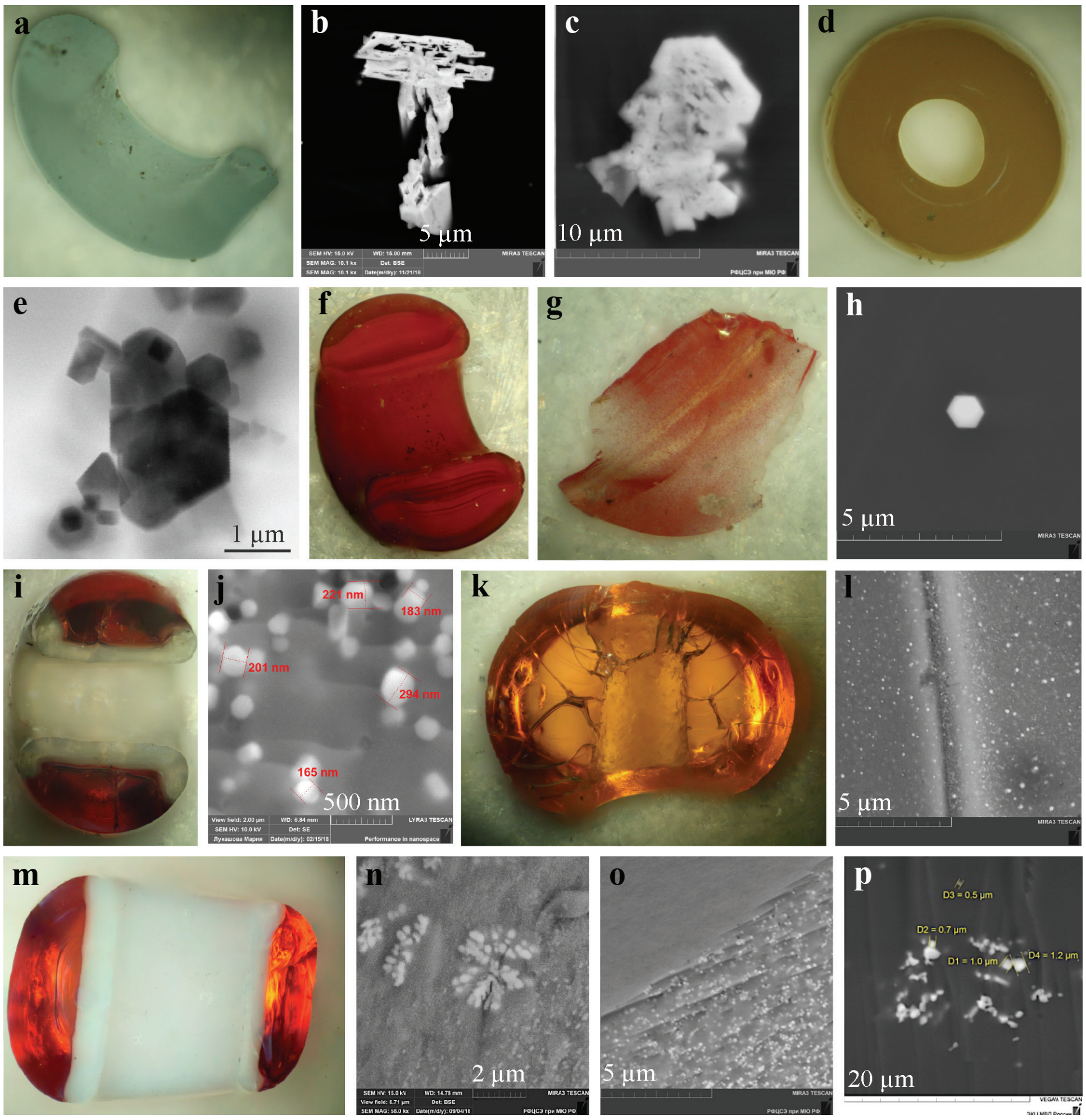

*Fig. 3.* SEM images of crystals detected in the studied historical glass beads:

(a) a fragment of the opal faceted bead (Sample 4) and (b, c) SEM images of cubic $NaSbO_3$ crystals;

(d) the yellow bead (Sample 5) and (e) a TEM image of cubic $Pb_2Fe_{0.5}Sb_{1.5}O_{6.5}$ crystals;

(f, g) fragments of the red opaque bead (Sample 7) and (h) a SEM image of an individual hexagonal CdZnSSe crystal;

(i) a fragment of the two-layered red-white bead (Sample 8) and (j) a SEM image of hexagonal $Ca_2Pb_3(AsO_4)_3Cl$ crystals in the white opaque layer;

(k) the peach bead (Sample 9) and (l) a SEM image of crystallites;

(m) the two-layered red-white bead (Sample 10) and (n) a SEM image of crystallites in the white opaque layer;

(o) a SEM image of crystallites in the white layer of a white-green bead;

(p) a SEM image of crystallites in a light-green bead.

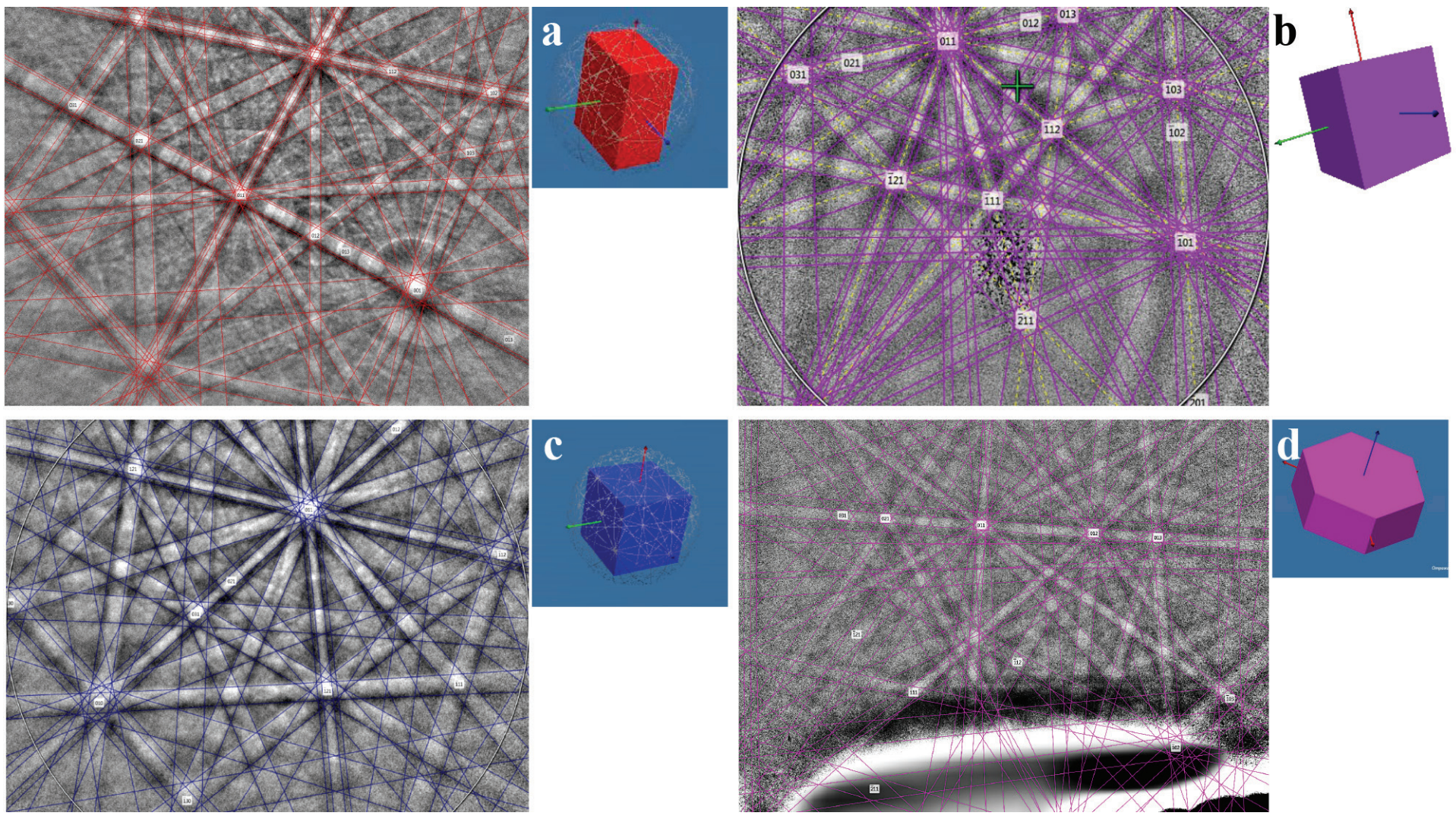

*Fig. 4.* EBSD patterns obtained from the precipitates of:

(a) orthorhombic $KSbOSiO_4$ in the turquoise bead (Sample 1);

(b) cubic $NaSbO_3$ in the opal faceted bead (Sample 4);

(c) cubic $Pb_2Fe_{0.5}Sb_{1.5}O_{6.5}$ in the yellow bead (Sample 5);

(d) hexagonal $Ca_2Pb_3(AsO_4)_3Cl$ in the white layer of the two-layered red-white bead (Sample 8).

*Table 1*. Elemental composition of the studied samples of 19th century glass beads.

| | | Elemental composition (at. %, EDX) | | | | | | | | | | | | | | | | | | Crystal structure (based on EDX + EBSD) |
|---|---|---|---|---|---|---|---|---|---|---|---|---|---|---|---|---|---|---|---|---|
| | | **Na** | **K** | **Si** | **Pb** | **S** | **P** | **Ca** | **Sb** | **As** | **Fe** | **Cd** | **Zn** | **Se** | **Cl** | **Cu** | **Al** | **Mg** | **Mn** | |
| **Sample 1,** *Fig. 2a, b* | Crystals | | 25.5 | 50.7 | | | | | 21.9 | | | | | | | 1.9 | | | | **Orthorombic $KSbOSiO_4$** |
| | Glass | 4.2 | 10.1 | 79.6 | 3.5 | | | | | | | | | | | 2.6 | | | | |
| **Sample 2** | Crystals + glass | 8.4 | 27.4 | 53.1 | 3.3 | | | 2.5 | 1.8 | | 0.2 | | | | 0.5 | 2.8 | | | | **Orthorombic $KSbOSiO_4$** |
| **Sample 3,** *Fig. 2c-e* | Crystals + glass (blue area) | 7.3 | 26.9 | 57.1 | 3.2 | | | 1.1 | 1.3 | | | | | | 1.0 | 2.1 | | | | **Orthorombic $KSbOSiO_4$** |
| | Crystals + glass (yellow area) | 2.6 | 12.9 | 72.8 | 4.0 | | | 1.8 | 2.5 | | | | | | 1.5 | 1.9 | | | | **Orthorombic $KSbOSiO_4$** |
| | Crystals + glass (black area) | 2.2 | 13.0 | 71.0 | 3.8 | | | 3.1 | 3.6 | | | | | | 1.0 | 2.3 | | | | **Orthorombic $KSbOSiO_4$** |
| **Sample 4,** *Fig. 3a-c* | Crystals | 13.3 | | 27.3 | | | | 20.8 | 38.6 | | | | | | | | | | | **Cubic $NaSbO_3$** |
| | Glass | 3.3 | 4.2 | 80.2 | | | | 7.7 | | | | | | | | | 4.5 | | | |
| **Sample 5,** *Fig. 3d, e* | Crystals | | | 50.9 | 28.7 | | | 5.3 | 12.8 | | 2.3 | | | | | | | | | **Cubic $Pb_2Fe_{0.5}Sb_{1.5}O_{6.5}$** |
| | Glass | | | 69.5 | 23.8 | | | 2.5 | | | | | | | | | 4.2 | | | |
| **Sample 6** | Glass | 7.2 | 1.9 | 65.0 | 20.2 | | | 3.1 | | | 1.1 | | | | | | | | 1.5 | |

| | | Elemental composition (at. %, EDX) | | | | | | | | | | | | | | | | | | Crystal structure (based on EDX + EBSD) |
|---|---|---|---|---|---|---|---|---|---|---|---|---|---|---|---|---|---|---|---|---|
| | | **Na** | **K** | **Si** | **Pb** | **S** | **P** | **Ca** | **Sb** | **As** | **Fe** | **Cd** | **Zn** | **Se** | **Cl** | **Cu** | **Al** | **Mg** | **Mn** | |
| **Sample 7,** *Fig. 3f-h* | Crystals | 15.0 | 4.1 | 40.9 | | 9.1 | | 0.8 | | | | 16.3 | 5.5 | 8.3 | | | | | | **Hexagonal CdZnSSe** |
| | Glass | 23.5 | 8.3 | 59.3 | | | | 1.5 | | | | | 7.4 | | | | | | | |
| **Sample 8,** *Fig. 3i, j* | Crystals in white glass | 4.4 | 2.0 | 12.1 | 28.1 | | 1.6 | 21.2 | | 22.7 | | | | | 8.0 | | | | | **Hexagonal $Ca_2Pb_3(AsO_4)_3Cl$** |
| | White glass | 2.6 | 8.9 | 70.2 | 9.0 | | | 2.3 | | 4.2 | | | | | 1.9 | | 1.0 | | | |
| | Red glass | 19.6 | 3.2 | 47.1 | | | 0.8 | 6.2 | | | 1.9 | | | | 1.0 | 15.3 | 1.8 | 3.3 | | |
| **Sample 9,** *Fig. 3k, l* | Crystals | 17.7 | 2.8 | 9.6 | 2.3 | | | | | 3.3 | | | 44.3 | | | | | | | **Unidentified** |
| | Glass | 3.9 | 8.9 | 76.0 | 5.2 | | | | | 6.0 | | | | | | | | | | |
| **Sample 11,** *Fig. 3o* | Crystals in white glass | 11.8 | 10.6 | 48.3 | 13.9 | | | 4.3 | | 9.1 | | | | | | 2.0 | | | | **Unidentified** |
| | White glass | 9.1 | 11.2 | 57.4 | 12.9 | | | 2.6 | | 4.3 | | | | | | 2.5 | | | | |
| | Green glass | 12.5 | 11.6 | 54.3 | 5.8 | | | 5.3 | | | 1.0 | | | | 0.9 | 6.5 | 0.8 | 1.3 | | |
| **Sample 12,** *Fig. 3p* | Crystals | 25.5 | 5.6 | 47.3 | | 6.8 | | 1.0 | | | | 6.8 | 7.0 | | | | | | | **Unidentified** |
| | Glass | 26.9 | 7.6 | 56.1 | | | | | | | | | 7.0 | | | | 2.4 | | | |